# Transcranial ultrasound simulations: a review

Authors: Célestine Angla[1,a], Benoit Larrat[2], Jean-Luc Gennisson[3], Sylvain Chatillon[1]

[1] Université Paris-Saclay, CEA, List, F-911120, Palaiseau, France

[2] Université Paris Saclay, CNRS, CEA, DRF/JOLIOT/NEUROSPIN/BAOBAB, Gif-sur-Yvette, France

[3] Université Paris Saclay, CNRS, Inserm, BioMaps, SHFJ, 4 Pl. du Général Leclerc, 91401 Orsay, France

[a] Author to whom correspondence should be addressed. Electronic mail: celestine.angla@cea.fr

Abstract

Transcranial ultrasound is more and more used for therapy and imaging of the brain. However, the skull is a highly attenuating and aberrating medium, with different structures and acoustic properties between samples and even within a sample. Thus, case-specific simulations are needed to perform transcranial focused ultrasound interventions safely. In this article, we provide a review of the different methods used to model the skull and to simulate ultrasound propagation through it.

0. Introduction

Transcranial ultrasound is more and more used for therapy and imaging of the brain. Therapeutic applications include thermal ablation [1], Blood-Brain-Barrier (BBB) opening [2] and neuro-modulation [3]. Transcranial ultrasound imaging has been used for a long time for vascular and flow imaging through the temporal windows but its generalization is still impeded by the strong image distortion and sensitivity loss at typical imaging frequencies [4].

Making ultrasound cross the skull is of high interest but it is also a big challenge as the skull is highly attenuating and aberrating. In addition, the pores located in the trabecular bone act as scatterers for ultrasound. As a result, ultrasound beams are both spatially shifted and attenuated. As the structure and acoustic properties of the skull differ between samples and even within a sample, case-specific simulations are needed to perform transcranial focused ultrasound interventions safely. The first issue when performing transcranial acoustic simulations is skull modelling. Due to the big variations within a skull and between specimens, most studies use images of the skull to extract its geometric description and derive acoustic properties using empirical relationships. However, these relationships are only hypothetical and the imaging resolution is often not precise enough to fully describe the porous structure of the skull. Once the skull is modelled, several simulation methods can be used, which can be divided into two main categories: numerical methods and semi-analytical ones. The type of method used usually depends on the application and the aim is to find the best balance between accuracy and computation performances. Likewise, approximations, such as neglecting shear waves and non-linear effects, can be made to speed up the simulations but the loss in accuracy has to be evaluated. After a brief overview of a few applications of transcranial acoustic simulations in the first part, parts one and two will focus on the different skull modelling methods that were previously introduced, and on the simulation methods used for transcranial ultrasound. Finally, the methods used to compare simulations and experiments will be described in the last part. Indeed, another main issue with transcranial acoustic simulations is to evaluate their accuracy. The only way to do so is to compare the results of the simulations with the results of experiments in a set-up as close as possible to the simulated one.

1. Applications of transcranial simulations

Transcranial acoustic simulations have many applications. Their main goal is to optimize the ultrasound treatment, by predicting the acoustic dose, the focalisation characteristics, and for operation planning. When combined with a multi-element transducer or an acoustic lens [5], simulations can be used to correct the phase and amplitude (especially at high frequencies where the distortion induced by the skull is higher) of the emitted signal in order to achieve good focusing through the skull or for imaging applications [6–8]. For thermal therapies, such as sub-thalamic nuclei ablation for the treatment of essential tremor [9], heat simulations are also often performed in order to predict whether heating occurs at the planned location and to avoid burning surrounding tissues. Acoustic simulations are thus needed in this case, as the pressure field obtained is the input of the thermal simulations. Thermal simulations are also used for non-thermal applications, in order to predict that no unwanted heating happens. Generally speaking, in addition to optimizing the treatment, acoustic simulations are often performed to ensure the safety of ultrasound therapies. Apart from heating prediction, they are used for standing waves and cavitation prediction [10–15] Finally, acoustic simulations can be used to optimize the positioning of the transducer [16].

### 1.1 Phase and amplitude correction

Transcranial simulations can be used to correct the phase difference induced by the skull in order to obtain nearly optimal focusing. To do that, several strategies can be used: classical time reversal method and cross correlation.

Time reversal consists in simulating wave propagation from a virtual source point located inside the brain at the intended focusing location and recording the signals on each element of the transducer. Then, if only time derivatives of even order are considered in the simulation equation, one can achieve focusing at the virtual source point by emitting the reversed signals received by each element of the transducer.

The cross correlation method aims to facilitate the experimental backpropagation, by using only the phase delays (and not the whole reversed signals). It is based on the following equation [17]:

$$p_{reference}.p_n(\tau) = \int p_{reference}(t)p_n(t+\tau)dt \quad (1)$$

Where $\tau$ is the phase delay.

As the skull is a strongly absorbing medium with varying thickness across its surface, one can find a way to distribute the emitted energy across the surface so the transducer elements near the thickest parts, and thus the most absorbing parts, of the skull, emit less power than the elements in front of the thinnest locations. Doing so will ensure a better transmission of the waves while avoiding unnecessary heating of the skull in the most absorbing locations. This distribution of energy can be achieved with amplitude correction [15,17–20].

Acoustic simulations can also be combined with the emission of a cavitation bubble that reflects the ultrasound waves [21]. This method uses acoustic simulations as a first step to focus near the intended location which enables the creation of a cavitation bubble, whose signal is recorded by the transducer array. The bubble is then used an invasive source inside the brain to obtain accurate focusing. For safety reasons, the relevance of this approach for clinical applications is debatable.

### 1.2 Heating prediction

The pressure field computed by the transcranial acoustic simulation can be entered as input to a heat transfer equation to simulate the heating of the skull and brain, and avoid burning tissues [19,20,22–28].

This heat transfer equation is called the Pennes Bio-Heat equation and is given by :

$$\rho C\frac{\partial T}{\partial t} = \nabla.(k\nabla T) - \rho_b w_b C_b(T - T_b) + \frac{\alpha p^2}{\rho c} \quad (1)$$

Where $\rho$ is the density of the tissue, $C$ its specific heat capacity, $k$ its thermal conductivity, $\alpha$ its absorption coefficient and $c$ its speed of sound. $\rho_b$ is the blood density, $C_b$ its specific heat capacity, $T_b$its temperature, and $\omega_b$ the perfusion rate. $p$ is ultrasound pressure derived from the acoustic model.

The first term of the right side models the thermal diffusion, the second one models the effects of perfusion and the last one models the acoustical power (found with the acoustic simulation). In Kyriakou et al. [20] and Guo et al. [26] add another term to the equation to model the body heat production. The Bio-Heat is often solved using 3D finite differences.

The criteria for a safe transcranial treatment are that the temperature at the focal region is higher than 60 °C and that there are no regions under the skull whose thermal dose is higher than 90 equivalent minutes at 43 °C [29].

Using the temperature found after solving the Bio-Heat equation, the thermal dose can be computed as:

$$TD = \int_0^t R\big(T(t')\big)^{T_{ref}-T(t')} dt' \quad (2)$$

With $T_{ref} = 43°C$ and $R(T) = \begin{cases} 0.50 & \text{for } \geq T_{ref} \\ 0.25 & T < T_{ref} \end{cases}$.

The thermal dose is defined as cumulative equivalent minutes at 43°C. For thermal ablation, the thermal dose can be used to compute the power required to achieve a thermal dose high enough to cause irreversible damage in the brain tissues. For instance, Pulkkinen et al. [27] perform the sonications with the power required to achieve an equivalent thermal dose of 25min at the focus. In addition to ensuring that the target location is heated enough to perform ablation, the thermal dose can also be used to check whether heating occurs at undesired locations. For example, in Pulkkinen et al. [27], target positions inducing a thermal dose of more than 5min at a location adjacent to the skull, were considered to be untreatable.

For non-thermal applications such as BBB opening, heat simulations can be performed to ensure that no undesired heating occurs during the treatment. For instance, Marquet et al. [22] estimate the maximum heat dissipation increase in the skull and find that after 60s of sonication, the temperature increase is lower than 0.03°C, which is negligible. In fact, this is due to the very short duty cycles, which let the skull bone cool enough between each pulse.

During most transcranial treatments, active cooling of the skull is performed in order to limit skull heating. Pulkkinen et al. [27] simulate the circulating water at 15°C for 15min prior to sonications using the Dirichlet boundary condition with constant temperature of 15°C around the skull. The simulated field is then used as the input temperature field for the sonications simulations. Kyriakou et al. [20] use a similar method to simulate the cooling procedure. Pulkkinen et al. [27] also simulate the active cooling of the skull base. This cooling method is based on circulating cooled water prior to sonications in a large nasal cavity at the centre of the skull-base. It is very important to apply such a method when targets are located near the skull base, as measurements have shown that the temperature in the soft tissue adjacent to the bone can exceed that of the focus [27]. In this study [27], the nasal cavity cooling was found to increase the treatment envelope for the uncorrected sonications, while for the phase corrected sonications, the nasal cavity cooling was found to sometimes have negative effect on the sonications if the focus was to close the cooled area (in those cases, the temperature also decreased at the focus).

Most thermal simulations are performed assuming that thermal tissue properties are not temperature dependant. However, the high temperature used for thermal ablation can cause vascular shutdown, which prevents the perfusion in those locations and cause the temperature to increase faster. Kyriakou et al. [20] simulate this effect by assuming that perfusion decreases linearly when the tissue temperature is above 50°C and stops above 51°C. The results show that after 20s of sonication, only 1 °C of the total additional temperature increase was observed in tissues where vascular shutdown occurred, suggesting that this effect can be neglected.

Heat simulations can also be used to predict thermal lesions. McDannold et al. [24] gathered the post-treatment images from 40 clinical TcMRgFUS treatments, including 16 for which bone marrow lesions were observed. They predicted the presence of lesions using a threshold for the acoustic energy of 18.1-21.1 kJ (maximum acoustic energy used) and 97.0-112.0 kJ (total acoustic energy applied over the whole treatment). The results show that the size of the lesions was not always predicted by the acoustic energy. However, the locations, sizes, and shapes of the heated regions estimated by the thermal simulations were qualitatively similar to those of the lesions and the lesions generally appeared in areas with high predicted temperatures.

### 1.3 Transcranial FUS safety assessment

Ultrasound can produce cavitation phenomenon and thus create bubbles that will oscillate with the varying pressure field. If the bubble oscillates too much and collapses, the pressure and temperature can increase very fast, and blood vessels can be damaged, causing haemorrhages. Simulations can be used to predict cavitation phenomenon [13–15] and avoid haemorrhages in the brain. Several studies have used simulation to try to understand phenomenon that occurred during clinical trials [13,30]. It can help avoid these effects in the future. However, simulation can also be used to prepare case-specific treatments and prevent those undesirable effects.

The standard indicator to evaluate the likelihood of cavitation risks is the mechanical index (MI), defined by:

$$MI = \frac{P_{-}}{\sqrt{f}} \tag{3}$$

With $P_{-}$ the peak rarefactional pressure and $f$ the frequency. This is only an indicator as cavitation depends on other factors that are currently unknown. To avoid adverse biological effects related to acoustic cavitation, the FDA (Food & Drug Administration - 510K Norm 1992) imposes the diagnostic devices to ensure a MI less than 1.9 [13].

In particular, studies [22] and [55] have focused on the TRUMBI (transcranial low-frequency ultrasound-mediated thrombolysis in brain ischemia) clinical trial which has been stopped prematurely because haemorrhages were observed on several patients. Baron et al. [13] simulate the pressure fields used in the trial, and they compute the associated MI. They find that the peak negative pressure is higher than the cavitation threshold in large areas of the brain, which is mainly due to the presence of standing waves. Using acoustic simulations, Pinton et al. [30] find that a volume of $2.7\text{cm}^3$ was above the MI threshold in the TRUMBI trial. They also investigate the risk of cavitation at 220kHz and 1MHz. They find that for an equivalent energy deposition rate and the same geometry, the brain volume above the MI ($0.4 \leq MI \leq 3$) is 3 to 4 orders of magnitude larger at 220kHz than at 1MHz. This high correlation between frequency and the volume with a high probability of cavitation is due to three reasons. First, the focal volume decreases with the frequency. Second, the heat deposition increases when frequency increases (as absorption increases). Third, the pressure needed to obtain a given MI decreases as the square root of the frequency. Both studies claim that unfocussed transducers (as the one used in the TRUMBI trial) should not be used, in order to avoid creating hot spots at unwanted locations due to the focusing effect of the skull.

Top et al. [15] use simulation to investigate blood-brain barrier disruption that was observed in the pre-focal region during previous experiments at 220kHz. Their results are in agreement with the disruption in the pre-focal region. However, they observe side lobes in the post-focal region in the simulations that did not seem to have had any effects in the experiments. They suggest this difference might be due to shielding of the ultrasound field due to microbubble activity in the focal region. They also simulate the field produced by the Passive Cavitation Detector (PCD) at its resonant frequency (610 kHz) and at the subharmonic of the transducer (110 kHz), as the sensitivity pattern is proportional to the transmitted field.

Transcranial acoustic simulations can also help predicting standing waves, which are contributors to undesired cavitation effects. Deffieux and Konofagou [10] and Mueller et al [11] estimate standing waves using a filter that detects their characteristic pattern. First, the fast oscillations of the field are extracted along each dimension using a high-pass spatial filter. Second, a Hilbert transform is applied to obtain the envelope of this signal. Third, only the maximum of the three dimensions is kept so as to have a unique map. The resulting field estimates the maximum of the spatial modulation pattern which is often associated with the presence of standing waves. Zhang et al. [12] use what they call the standing-wave ratio to predict the intensity of standing waves. It is defined as:

$$R_a = \frac{P_{max} - P_{min}}{P_{avg}} \tag{4}$$

Methods for reducing standing waves have also been developed and verified thanks to simulations. Baron et al. [13] suggest using higher frequencies (more absorption so less interferences), reducing the duration of the pulses (to avoid interferences) and applying modulation. Deffieux and Konofagou [10] propose to use fast periodic linear chirps to reduce standing waves. Chirps are signals with a varying frequency over time. It induces a time-dependent phase difference between the incident and reflected waves which ensures that, even if interferences occur, their position will change over time and thus, after some time, the effects of constructive and destructive interference will even out. When using a chirp with a period of 23µs and a frequency oscillating between 450 and 550 kHz, they manage to reduce the peak amplitude of standing waves (as a percentage of the peak pressure) from 19% to 12%, and to reduce the standing wave volume (that is to say the volume with standing wave amplitude higher than 5% of the peak pressure) from 0.87% to 0.17%. While the standing wave volume reduction is quite satisfying (5 times reduction), the standing wave amplitude reduction is not as significant (1.5 times reduction). The authors suggest it may be due to interferences occurring close to the skull surface, because the time delay between the incident and reflected waves is too small compared to the frequency shift. Zhang et al. [12] use the Randi function to generate a random phase between $0$ and $\pi$, which is added to the original signal at regular intervals to break the standing wave formation condition.

## 2. Skull modelling

Skull modelling is an important factor for accurate simulations. As the shape and acoustical properties of the skull vary between individuals, and even within one individual (across the years and among the whole surface of the skull), it is very tricky to create good skull models. Most skull models are based on skull imaging of the considered individual, as for example X-rays.

### 2.1 Shape

When finite differences or other voxel-based simulation algorithms are used, the skull shape does not need to be extracted. However, for geometrical based simulation algorithm, one needs to model the skull shape.

Hayner and Hynynen [31] model the skull as a medium contained between two flat, but not necessarily parallel, interfaces, as in the frequency range considered, the minimum radius of curvature of a skull is greater than the wavelength.

However, in most studies, the skull is segmented from either computed tomography (CT) scans or magnetic resonance (MR) images, in order to have a more accurate modelling of the geometry. To obtain the shape of the skull from a tensor of intensity voxels, two steps are usually needed: a segmentation step and a mesh-making step.

Segmentation of the bone voxels is often performed using a threshold method. Clement and Hynynen [32], Clement et al. [33] and Yin and Hynynen [34] identify the inner and outer surfaces of the skull by searching for the innermost and outermost densities greater than $1400\ kg.m^{-3}$ along each line of a CT slice. Pichardo and Hynynen [35] perform the segmentation with the FMRIB Software Library [36].

Mesh-making is often performed using an iso-surface algorithm [35,37–40]. In some cases, a condition is imposed so that no surface area element is greater than $\left(\frac{\lambda}{2}\right)^2$, with $\lambda$ the acoustic wavelength in water [38,39].

For skull segmentation from MRI, two UTE (ultra-short echo time) images are generally combined (using simple arithmetic operations) to separate skull voxels from other tissues using various thresholds [26,41]. In Miller et al. [41], after thresholding, a spatial connectivity requirement is applied to the pixels in the bone class, to eliminate isolated pixels or groups of pixels that are clearly not part of the skull.

### 2.2 Acoustical properties

Several studies have measured skull acoustical properties (density, speed of sound and attenuation) [31,42,43]. Even though properties of pure bone could be tabulated, most of the bone volume is heterogeneous and it is hard to predict the effective properties due to partial volume effect of skull images. However, most studies try to deduce the acoustical properties of a given skull from images, acquired with either CT or MR.

### 2.2.1 Density

#### 2.2.1.1 Linear relationship between density and Hounsfield Units (HU)

In Rho et al. [44], ultrasonic measurements show a linear relationship between density and HU. HU are arbitrary units that depend on the CT scanner used. It is thus necessary to have a specific conversion from HU to any physical measure for each sample.

Connor et al. [45] introduce a method to compute the skull density from HU. Assuming that the relation between HU and density is linear, and ensuring that a sample of both water and air (which have known densities) appear in the CT scan, one has:

$$\rho = \kappa_1 H + \kappa_0 \tag{6}$$

With $\kappa_1 = \frac{1}{H_{water} - H_{air}}$ and $\kappa_0 = \frac{-H_{air}}{H_{water} - H_{air}}$.

This method is used in [6,7,28,37–39,46–48].

To confirm the linear relationship obtained, they make the hypothesis that density and HU are related by a second order polynomial. They use the relationship relating the total mass of the skull M with the volume $v$ of a voxel and the sum of all densities across the skull: $M = v \sum_{skull} \rho$. Using this additional relationship, they are able to fit a second order polynomial for density, but they find that the quadratic term is negligible in front of the others (by five orders of magnitude). Thus, their linear approximation seems correct.

#### 2.2.1.2 Porosity-based and equivalent relationships

Aubry et al. [18] propose a different method based on the bone porosity. HU are defined by:

$$H = 1000 \frac{\mu_x - \mu_{water}}{\mu_{bone} - \mu_{water}} \tag{7}$$

where $\mu_x$, $\mu_{water}$ and $\mu_{bone}$ denote the photoelectric linear attenuation of the explored tissue, water and bone. They propose a linear relationship:

$$\mu_x = \Psi \, \mu_{water} + (1 - \Psi)\mu_{bone} \tag{8}$$

with $\Psi$ the bone porosity. Thus they have:

$$\psi = 1 - \frac{H}{1000} \tag{9}$$

The density can be computed from the porosity with:

$$\rho = \psi\rho_{water} + (1 - \psi)\rho_{bone} \tag{10}$$

This method is used in [11,12,52,13,15,19–21,49–51].

This relationship implies a linear relationship between the bone density and the HU, as it was assumed in Section 1.2.1.1. However, instead of building the linear fit from $H_{water}$ and $H_{air}$, they build it from $\rho_{water}$ and $\rho_{bone}$. In the first case, $H_{water}$ and $H_{air}$ are derived from direct measurements on the CT scans. Thus, they are more reliable than the values of $\rho_{water}$ and $\rho_{bone}$, which are taken from measurements from the literature that were probably not conducted in the exact same conditions as the experiments conducted in the study. On the other hand, if the density is assumed to be linearly related to HU in the range corresponding to bone, it is probably not the case for all values of HU. Thus, building the linear fit from values in water and cortical bone (which are the lower and upper bounds for HU in the skull) seems more accurate than using values from air and water.

Guo et al. [26] use the same equations as before, but they change the definition of porosity:

$$\psi = 1 - \frac{H}{\max(H)} \tag{11}$$

Indeed, with Aubry's definition of Hounsfield Units, one has: $0 \leq H \leq 1000$. In this study, they generalize Aubry's formulas for raw CT data, where the maximum value of $H$ is not necessarily $1000$.

Deffieux and Konofagou [10] normalize the apparent CT density to have $0 \leq \rho_{CT} \leq 1$, and then the density is given by:

$$\rho = \rho_{min} + (\rho_{max} - \rho_{min})\rho_{CT} \tag{12}$$

This is in fact equivalent to the previous equation based on porosity as $\rho_{CT} = \frac{H}{\max(H)} = 1 - \Psi$.

Marsac et al. [53] introduce a new relationship for density:

$$\rho = \rho_{min} + (\rho_{max} - \rho_{min})\frac{H - H_{min}}{H_{max} - H_{min}} \tag{13}$$

This equation is similar to the porosity-based equation (10), and is equivalent to it when $H_{min} = 0$, Indeed, if $H_{min} = 0$, one has: $1 - \Psi = \frac{H - H_{min}}{H_{max} - H_{min}}$. Thus this equation is the more general form of the porosity based equation. This equation is used in [5,54].

#### 2.2.1.3 Bone fraction based relationship

Vyas et al. [55] propose a method based on the fraction of bone in the voxel, which is defined by:

$$f = \frac{\frac{H}{1000} \times \frac{\mu}{\rho}_{water_eff}}{\rho_{bone} \times \frac{\mu}{\rho}_{bone_eff} - \frac{\mu}{\rho}_{water_eff}} \tag{14}$$

With $\frac{\mu}{\rho}_{eff} = \frac{\Sigma S(E) \times \frac{\mu}{\rho}(E)_{water}}{\Sigma S(E)}$, where $S(E)$ is the beam spectrum as a function of energy. Then the density is defined by:

$$\rho = f \times \rho_{bone} + (1 - f)\, \rho_{water} \tag{15}$$

This method is similar to the porosity-based equation (10). However, bone fraction is computed in a more complicated way than porosity, and equation (14) seems not to be homogeneous.

### 2.2.2 Speed of sound

#### 2.2.2.1 Linear relationships between speed of sound and Hounsfield Units

Fry and Barger [56] and in Vyas et al [55] use an empirical linear relationship linking directly the speed of sound to Hounsfield Units:

$$c = 1460 + 0.7096\, H \tag{16}$$

Clement and Hynynen [32] use a relationship between the average speed of sound in a skull and the average density of that same skull, obtained experimentally from 1000 measurements with 10 skulls at 0.51MHz:

$$c = 2.06\, \rho - 1540 \tag{17}$$

With $c$ in $m.s^{-1}$, for densities $\rho$ between $1820 kg.m^{-3}$ and $2450 kg.m^{-3}$. Thus $c \in [2209;\ 3507]\ \mathrm{m.s}^{-1}$. This roughly corresponds to the range of values found in the literature.

As a linear relationship is assumed between density and HU, this relationship is similar to equation (16) as it implies a linear relationship between speed of sound and HU.

#### 2.2.2.2 Porosity-based and equivalent relationships

Similarly to the porosity based relationship for density, Aubry et al. [18] propose a sound speed relationship based on porosity:

$$c = c_{min} + (1 - \psi)(c_{max} - c_{min}) \tag{19}$$

The linear relationship between velocity and porosity is justified by Carter and Hayes who showed that the elastic modulus of bone is proportional to the apparent density cubed. This method is used in [11,12,51,52,13,15,19–21,26,49,50]. This relationship is similar to equations (16) and (17) as it implies a linear fit between speed of sound and HU. However, equation (19) is more general as porosity does not depend on the CT scanner while HU do.

In the same way as for the density equations, other studies have improved the porosity based equation for speed of sound so as to make it more general. In order to account for CT scans where HU are not normalized, Deffieux and Konofagou [10] propose:

$$c = c_{min} + (c_{max} - c_{min})\rho_{CT} \tag{20}$$

Which is equivalent to the equation introduced by Marsac et al. [53]:

$$c = c_{min} + (c_{max} - c_{min}) \frac{\rho - \rho_{min}}{\rho_{max} - \rho_{min}} \tag{21}$$

These equations are also used in [5,54].

Marsac et al. [53] search for the best value of $c_{max}$ such that the simulations fit the experiments.

#### 2.2.2.3 Relationships obtained with genetic algorithms

Connor et al. [45] use a genetic algorithm to find the optimal relationship between density and speed of sound at 0.74MHz, using a success function that compares the phase difference between simulation and experiment.

They compare their result function with what could be expected for a cellular solid, as they approximate the skull bone as one. The speed of sound in a solid is given by $c = \sqrt{\frac{E}{\rho}}$ where E is the Young's modulus of the material and ρ its density. For open-pore cellular materials it can be shown that $E \propto \rho^2$ while for closed-pore materials $E \propto \rho^3$. The curve of speed of sound against bone density obtained in this study agrees with these models. Indeed, it initially has a form similar to that of a square root function (as it would be for an open-celled porous solid), then the model transitions into a linear function (as it would be for a closed-celled porous solid). In addition, the transitional density region is located between the density of trabecular bone and that of cortical bone.

Pichardo et al. [37] use a similar method, but they investigate the sound speed relationships at various frequencies as they have found from measurements that the skull is a dispersive medium. They perform a two-step optimization (to speed up the process while maintaining a good simulation accuracy) at several frequencies: 0.27, 0.836, 1.402, 1.965 and 2.525MHz. The sound speed functions found in this study were used (either directly or interpolated to fit the needed frequency) in [6,7,28,38,39,46,47].

#### 2.2.2.4 Polynomial relationship

McDannold et al. [23,24] assume that the relationship between the inverse of the speed of sound and the skull density can be approximated by a series of polynomials:

$$\frac{1}{c} = \sum_{m} B_m \rho^m \tag{5}$$

The advantage of this formulation is that they can estimate the phase shifts resulting from a change in speed of sound with:

$$\Delta\phi_k(x_i, y_i) \approx 2\,\pi f \sum_{j=0}^{i} \frac{\Delta z}{c(x_i, y_i, z_j)} = 2\,\pi f\ \Delta z \sum_{m} B_m \sum_{j=0}^{i} \rho^m(x_i, y_i, z_j) \tag{23}$$

Indeed, when thermal therapy is performed with ultrasound, they observe changes in the speed of sound inside the skull due to temperature changes, thus this formula allows adapting the phase shift during therapy. However, this polynomial relationship seems to go against all previous models that proposed a relationship proportional to density.

### 2.2.3 Attenuation

#### 2.2.3.1 Constant attenuation

Deffieux and Konofagou [10] consider a constant attenuation across the whole skull as they claim that more complex models for attenuation can be inconsistent.

#### 2.2.3.2 Porosity based and normalized density based relationships

Similarly to their density and sound speed relationships, Aubry et al. [18] propose a bone porosity based relationship for attenuation:

$$\alpha = \alpha_{min} + (\alpha_{max} - \alpha_{min})\psi^{\beta} \tag{24}$$

The constants were adjusted by comparing simulations with measurements. This method is used in [11,12,52,13,15,19–21,49–51].

Yoon et al. [57] use a method based on the normalized density, as was done by Deffieux and Konofagou [10] for density and speed of sound:

$$\alpha = \alpha_{min} + (\alpha_{max} - \alpha_{min})\rho_{CT} \tag{25}$$

Unlike the relationships for density and speed of sound proposed with the normalized apparent density, this formula is not equivalent to the porosity based attenuation formula. Indeed, this is a linear relationship and not a power law, and $\rho_{CT} = \frac{H}{\max(H)} = 1 - \Psi \neq \Psi$.

#### 2.2.3.3 Bone fraction based relationship

Vyas et al. [55] propose a method based on the fraction of bone in the voxel, which is defined by:

$$f = \frac{\frac{H}{1000} \times \frac{\mu}{\rho}_{water_eff}}{\rho_{bone} \times \frac{\mu}{\rho}_{bone_eff} - \frac{\mu}{\rho}_{water_eff}} \tag{26}$$

With $\frac{\mu}{\rho}_{eff} = \frac{\Sigma S(E) \times \frac{\mu}{\rho}(E)_{water}}{\Sigma S(E)}$, where $S(E)$ is the beam spectrum as a function of energy. Then the absorption component of attenuation is defined by:

$$\alpha_a = f \times \alpha_{bone} + (1 - f)\,\alpha_{soft\ tissue} \tag{27}$$

For the scattering component of the attenuation they use data found experimentally by Tavakoli [58]. This data consists in experimental values of attenuation for various porosity values. They fit the data to linear curves and obtain four linear equations (for four ranges of porosity) relating the scattering component of attenuation with porosity. Then they use those equations to compute the scattering component of attenuation, and they use the voxel bone fraction to compute porosity:

$$\Psi = \frac{f_{maxsubject} - f}{f_{maxsubject}} + 0.05 \tag{28}$$

#### 2.2.3.4 Relationships obtained with genetic algorithms

Pichardo et al. [37] use a genetic algorithm to find a relationship between attenuation and density at several frequencies : 0.27, 0.836, 1.402, 1.965 and 2.525MHz. Like in 1.2.2.3, they find the relationship that minimizes the phase difference between simulation and experiment. The attenuation functions found in this study were used (either directly or interpolated to fit the needed frequency) in [6,7,28,38,39,46,47].

#### 2.2.3.5 Polynomial relationship

McDannold et al. [23,24] assume that the relationship between the attenuation and the skull density can be approximated by a series of polynomials:

$$\alpha = \sum_{m} A_m \rho^m \tag{29}$$

The advantage of such a formulation is that, assuming that attenuation mainly occurs along the propagation axis $z$, one can compute the pressure field without attenuation and deduce an approximation of the pressure field with a given attenuation model without having to run again the whole simulation. Indeed, if attenuation mainly occurs along the $z$ direction, the pressure distribution for element $k$ is approximately:

$$P_k(x_i, y_i, z_i) \approx P_{k0}(x_i, y_i, z_i) \times \exp\left(-\sum_{j=0}^{i} \alpha(x_i, y_i, z_j)\ \Delta z\right) \tag{30}$$

Where $P_{k0}$is the pressure computed without attenuation for the transducer element number $k$. This formula shows that the pressure field without attenuation can be computed separately from the attenuation contribution with this method, allowing to compare several attenuation models without any loss of computation time.

#### 2.2.3.6 Frequency dependency of attenuation

In Pichardo et al. [37], the relationships between density and attenuation found at several frequencies suggest that attenuation generally increases with frequency. Several studies have focused on the relationship between attenuation and frequency.

##### 2.2.3.6.1 Linear relationships

Connor et al. [45] use linear increase of attenuation with frequency with various coefficients, one for cortical bone:

$$\alpha = 167 \times f\ .10^{-6}\ Np.m^{-1} \tag{31}$$

And one for trabecular bone:

$$\alpha = 300 \times f\ .10^{-6}\ Np.m^{-1} \tag{32}$$

Where trabecular bone voxels are identified as those containing less than 70% of bone.

Bossy et al. [59] perform acoustic simulations through trabecular bone at 0.4-1.2MHz. They find that the attenuation linearly increases with frequency and can thus be modelled by:

$$\alpha = \alpha_0 + nBUA \times f \tag{33}$$

They also find that nBUA values are strongly and positively correlated with the bone volume fraction (BV/TV) and that speed of sound is positively correlated with the bone volume fraction.

Haïat et al. [60] confirm that both Broadband Ultrasonic Attenuation (BUA) and speed of sound exhibit a strong positive correlation with the bone volume fraction. In addition, they claim that BUA and speed of sound vary quasi-linearly with the bone volume fraction.

###### 2.2.3.6.2 Power laws

Attenuation is a combination of absorption and scattering. According to Pinton et al. [14], 86% of attenuation is due to scattering (and thus only 14% is due to absorption) in skull bone at 1MHz. Thus, a more efficient model of attenuation could be defined, by separating the effects of absorption and scattering. That is what is done in Yousefian et al. [61], where they numerically study attenuation in porous media (mimicking cortical bone). They claim that total attenuation is not described by a linear combination of scattering and absorption anymore in the presence of multiple scattering. Thus, they propose a non-linear formula for total attenuation:

$$\alpha_{tot} = \alpha_{scat} + c\, f^{\beta_{app}} \tag{34}$$

Where $\beta_{app}$ and c are to be determined and depend on pore diameter and pore density.

Webb et al. [62] compare CT and MR images as predictors of attenuation. They measure the acoustic attenuation at 0.5, 1, and 2.25 MHz in 89 samples taken from two ex-vivo human skulls and find the best parameters to fit the equation:

$$\alpha = \alpha_0 f^{\beta} e^{cp} \tag{35}$$

Where $f$ is the frequency, $p$ the imaging parameter (HU value, UTE and ZTE magnitude, or $T_{2^*}$ value), and $\alpha_0$, $\beta$, $c$ the model parameters. This equation assumes the traditional power law between frequency and attenuation, and that the imaging parameter provides a rough estimate of the pore structure in each sample. HU provide the best prediction of attenuation with a minimum standard error of 1.7 Np/cm. The ZTE, UTE, and $T_{2^*}$ values had standard errors of 2.0, 2.0, and 2.1 respectively.

### 2.2.4 Deriving acoustical properties from MRI

While acoustical properties are most often derived from CT-scans, several studies have investigated the possibility of using MR images (as MR is often used to guide the focused ultrasound therapy), in order to reduce the whole procedure and to avoid the patient exposure to X-rays. These methods consist in building a virtual CT data from MRI.

Wintermark et al. [63] compare three MRI pulse sequences: $T_1$ weighted 3D Volumetric Interpolated Breath-hold Examination (VIBE), proton density weighted 3D Sampling Perfection with Application-optimized Contrasts using different flip-angle Evolution (SPACE) and 3D true Fast Imaging with Steady-state Precession (FISP) $T_2$-weighted imaging). The MR modality giving a total thickness the closest to the CT-based total thickness was identified. Random coefficient regression was used to predict CT total skull thickness based on the optimal MRI sequence.

The same method was used for the thickness of each of the three layers (inner table, diploe, outer table). Similarly, a regression model was used to predict reference standard CT average density based on the optimal MRI sequence average intensity. Virtual CT datasets derived from the MRI datasets were built using the models described before. The $T_1$ 3D VIBE sequence was the MRI sequence coming closest to the reference standard CT in terms of measuring total skull thickness, and was thus selected for the rest. In experiments made with a human skull, the mean absolute difference between the phase shifts calculated with standard CT and virtual CT was 0.8 ± 0.6 rad.

Miller et al. [41] investigate the feasibility of using ultrashort echo-time magnetic resonance imaging (UTE MRI) instead of CT to calculate and apply aberration corrections on a clinical TcMRgFUS system. A 3D map of the skull is created from the MR images by applying a segmentation algorithm, and is used to construct virtual CT data, by assigning 1000 Hounsfield units to bone pixels and −1000 Hounsfield units to all other pixels. The virtual CT data was entered in the planning software of a MRgFUS system to calculate aberration correction phases. There was no significant difference between the sonication results achieved using CT-based and MR-based aberration correction.

#### 2.2.5 Shear waves parameters

As the acoustical parameters of shear waves in bone have not been studied a lot, most studies that account for the presence of shear waves assume that the shear waves speed is half the one of the longitudinal waves, as it is a typical assumption for solids. However, a few studies have looked more in details into shear wave parameters.

White et al. [42] measure an average speed of sound for a 1.0 MHz longitudinal wave of 2820±40 m/s and 1500±140 m/s for shear waves. They find that the shear attenuation coefficient in skull bone is on average higher by 115 Np/m than the longitudinal one for the frequency range studied. So, although the speed of sound of shear waves in the skull is closer to the one of the surrounding media (brain and tissues) which allows a more efficient transmission through the boundaries, the high shear attenuation coefficient tends to restrict this beneficial effect.

Pichardo et al. [48] use a genetic algorithm method (as they did before [37] for compressional waves parameters) to establish the relationship between apparent density calculated from CT scans and shear speed of sound and attenuation at 270 kHz and 836 kHz. They assume a linear relationship for speed of sound and a constant attenuation.

#### 2.2.6 Comparison of the methods and discussion

##### 2.2.6.1 Comparison of a few methods for acoustical parameters modelling

McDannold. et al [23] compare Pichardo relationships [37] for the speed of sound and attenuation to their polynomial formulas at 660kHz. While there was no significant difference between their relationships and Pichardo's in the resulting dimensions or obliquity of the simulated focal region, the temperatures simulated using Pichardo relationships were less accurate (less close to the experiments) than those simulated using their relationships.

Bancel et al. [54] use Marsac density relationship [53] and compare the speed of sound relationships given by Pichardo [37], Marsac [53] and McDannold [23]. Marsac relationships seem to perform a bit better for aberration correction simulations, with a restored pressure, compared to the hydrophone-based correction, of 85% (compared to 81% with Pichardo relationship and 82.5% with McDannold relationship) and a peak to side lobe ratio of 45.5% (compared to 51.5% with Pichardo or McDannold relationship).

#### 2.2.6.2 Impact of the error in skull modelling

In Robertson et al. paper [49], the impact of changes in bone layer geometry and the speed of sound, density, and acoustic absorption values is quantified through a numerical sensitivity analysis. The errors in the field resulting from noisy variations (around a mean value) in medium properties are smaller than those from linear changes (variations of the mean value). Noisy variations for absorption has almost no effect: homogeneous maps of absorption are good. Linear variations in HU values resulted in errors lower than the errors for individual changes in assigned medium properties, thus suggesting that the primary concern should be the robustness of any method for the conversion of CT images to acoustic properties. Speed of sound is shown to be the most influential acoustic property, and must be defined with less than 4% error. Changes in the skull thickness of as little as 0.1mm can cause an error in peak pressure of greater than 5%, while smoothing with a $1\ mm^3$ kernel (skull maps obtained from low resolution images) causes an increase of over 50% in peak pressure.

#### 2.2.6.3 Considering X-ray energy

While many studies have investigated the relationship between acoustical properties and Hounsfield Units, they have ignored the impact of X-ray energy (HU is a function of the linear attenuation coefficient which itself depends on the X-ray energy), reconstruction method, and reconstruction kernel on the measured HU. Webb et al. [64] imaged 91 human skull fragments by 80 CT scans with a variety of energies and reconstruction methods. The average HU from each fragment is found for each scan and correlated with the speed of sound measured. The results show that both the energy and the reconstruction method have a significant influence on the relationship between velocity and HU. The main issue is that it is difficult to have an estimate of the real X-ray energy except in dual-energy CT. Aubry and Marsac relationships are based on porosity, which depends on the linear attenuation and thus accounts for X-ray energy. So these formulas work well when the energy is known. Pichardo and Connor relationships do not account for X-ray energy. Considering CT parameters can be obtained with dual-energy CT, by obtaining calibration measurements or by working with the vendor to obtain accurate estimates of the X-ray spectrum of a given scan. The results provide estimates of the relationship between HU and velocity for a variety of different CT parameters. They also show that the relationship between HU and velocity is patient-specific. The measurements show that CT is able to capture only about one-half of the variation in acoustic velocity within the skull. Some of the remaining variation is likely due to errors in the measured velocity, but it is also likely an indication that some of the variation in velocity is not well modelled by HUs. This could be because the variation in velocity results from changes in the chemical composition of the skull, a variation not necessarily captured by the measured HU.

### 2.3 Level of heterogeneity

The skull is made of two types of bone: cortical bone and trabecular bone. The inner and outer tables are made of cortical bone, which is a dense and nearly homogeneous medium, whereas the middle layer is made of trabecular bone, which is a complex porous structure, highly scattering. Thus, the skull is very heterogeneous. However the imaging modalities, be it CT or MR, cannot capture heterogeneities below their resolution. Thus, when modelling the skull, one can use levels of heterogeneity going from assigning different acoustical properties to each voxel, to assigning the same acoustical properties to the whole skull.

#### 2.3.1 Fully heterogeneous

In this model, each voxel has a given density, sound speed and attenuation. It is used in [5,6,24–26,28,37,45–47,49,50,10,51–53,57,65–68,13,15,18–21,23].

#### 2.3.2 Heterogeneous but binary

The skull is assumed to be a complex porous structure made of bone and with pores filled with marrow (whose acoustical properties are similar to those of water). Simple image thresholding allows labbelling voxels as either being bone or marrow and constant acoustical properties are assigned to bone or marrow. It is used in [14,17,30,59,60,69]. However, this method requires high resolution scans, as the structure of the diploe is not visible at the millimeter scale.

#### 2.3.3 N layers

In this model [37], the skull is divided into N homogeneous layers. Density along the acoustic axis (only normal incidence is considered) was computed by averaging the density from the voxel on the acoustic axis and the one of its four neighbours in the plane perpendicular to the acoustic axis. In a given layer, the acoustical properties are averaged from the CT scans. Each layer thickness is set to ¼ of the voxel spacing to allow a smooth variation.

#### 2.3.4 Three layers

The skull is often described as a three-layer medium representing the inner and outer tables made of cortical bone and the middle layer made of trabecular bone. In each layer, the acoustical properties are averaged from the CT scans [31,63].

#### 2.3.5 One layer

The simplest assumption that can be made is considering the skull as a homogenous and isotropic medium. The acoustical properties are either averaged from CT scans, or taken from measurements from the literature. This model is used in [6,20,32–35,41,49,52].

In Jiang et al. [52], the velocity on the skull surface is taken to be that in cortical bone (which influences refraction computations) and the velocity inside the skull is taken as being the averaged value on the ray paths (which influences the time of flight computations). The density is taken as being the maximum density as it influences the refraction computations.

The main issue with those methods is that if the transducer has a large aperture, the acoustical properties may vary quite a lot along the surface intersected by the ultrasound beam. In Jones et al. [39] and Pajek and Hynynen [38], a locally averaging method was proposed. It consists in spatially averaging the acoustical properties independently for each ray path going through the skull.

#### 2.3.6 Comparison of the levels of heterogeneity

Jones and Hynynen [6] experimentally compare CT-based aberration corrections with two skull models : one homogeneous layer and a fully heterogeneous model. For both models, they compare the corrections computed using a full-wave method and using a ray-tracing method. By averaging the metrics on the two methods, the homogeneous model increases the peak pressure shift error (compared to the hydrophone-based shift) by 4%, the -3dB volume error by 27.5% and the peak-to-side-lobe ratio error by 22%, compared to the heterogeneous model. Thus, the heterogeneous model performs better, as the homogeneous one underestimate attenuation and the focal spot spreading.

Kyriakou et al. [20] find similar results when comparing heterogeneous and homogeneous skull modelling with a full-wave simulation. Indeed, the prediction of the focusing (shift and volume) seems unaffected by skull inhomogeneity while the peak pressure is reduced by 25% with the heterogeneous model.

Jiang et al. [52] compare the performances of two homogenous models : one with the average properties of the skull, and one the properties of the cortical bone on the surface (as refraction computations are influenced only by the properties on the surface on the skull) and with the average properties of the skull on the inside (for the time-of-flight computations). Shifts of 9.5 mm in the axial direction and 1.5 mm in the focal plane compared with the virtual source were observed with the first one, compared to 0.5 mm in the axial direction and 0.5 mm in the focal plane with the optimized model.

Robertson et al. [69] study the influence of homogenization of CT maps. Progressive homogenization of acoustic property maps leads to an overestimation in the amplitude of transmitted US, an underestimation of time-of-flight, and a loss of fine spatial detail in the intracranial field. Inflating the simulated attenuation coefficient of the skull layer reduces the error in transmitted pressure amplitude to around 40%, however this is unable to correct fully for errors in time of flight and the pressure distribution of the transmitted field.

One must also take into account that the frequency will influence the level on heterogeneity needed. Indeed, at lower frequencies, heterogeneities much smaller than the wavelength do not need to be taken into account and parameters at those scales can be homogenised.

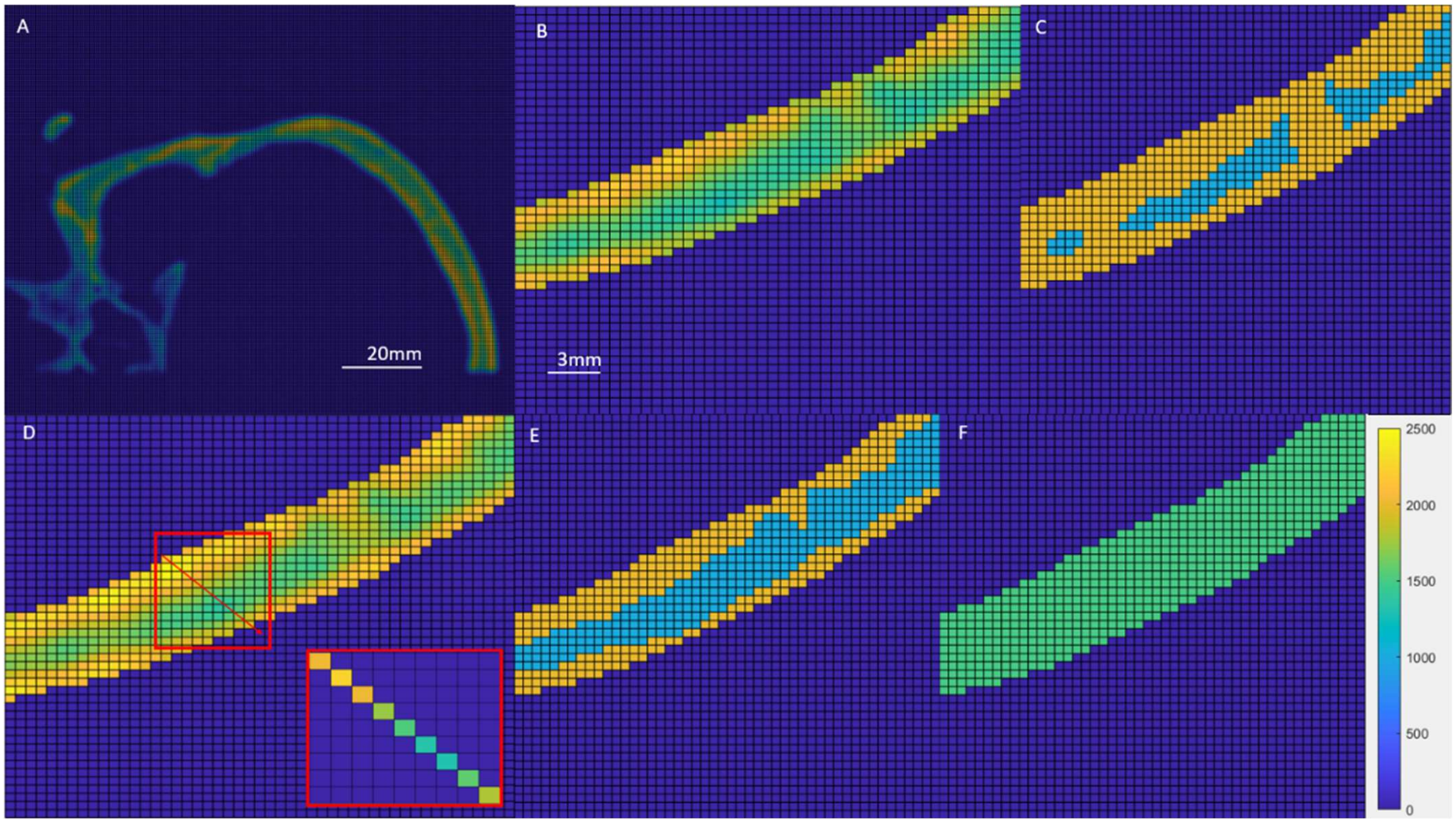

*Figure 1 : The different level of homogenization. A: original skull, B: heterogeneous model, C: binary model, D: N-layers model, E: three-layers model, F: one layer model).*

3. Simulation methods

3.1 Semi-analytical methods

3.1.1 Ray tracing

A fast common method to simulate wave propagation is ray tracing [33–35,38,39,42,52]. It considers reflection, refraction and mode conversion at each interface. The ray paths are obtained using Snell-Descartes' laws of a three-layer model, composed of water, skull and brain.

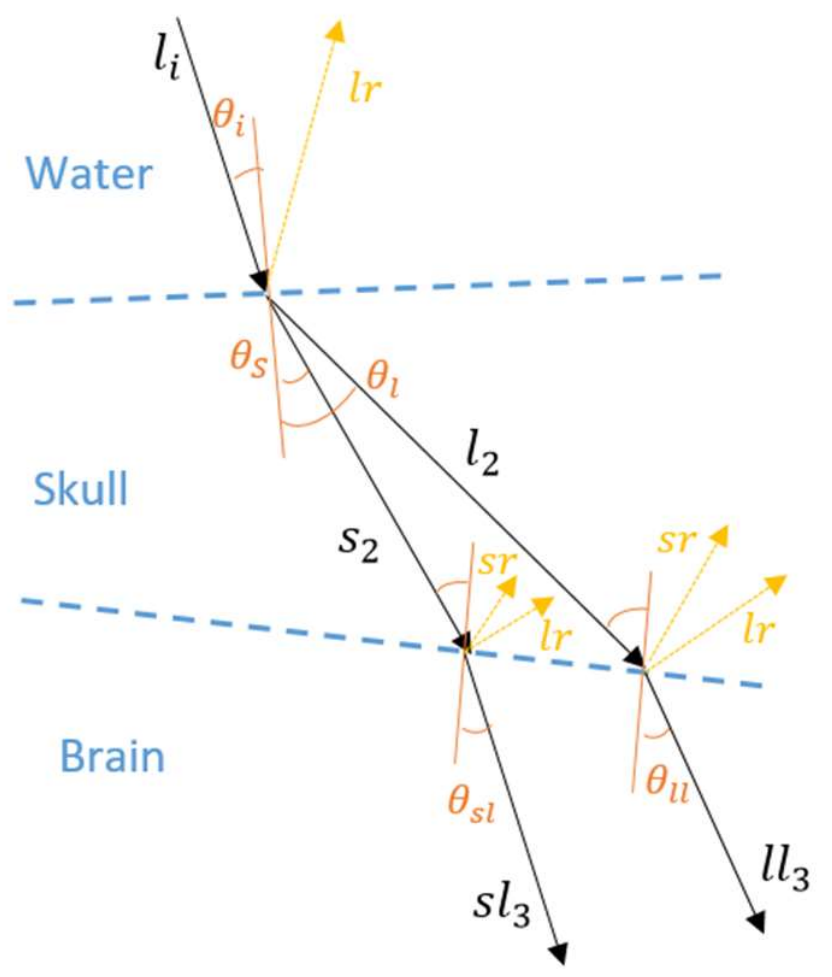

*Figure 2: three-layer model of wave of transmission through the skull*

In general, possible reflections from inside the skull are not considered due to the high attenuation of the skull [35]. Moreover, as the skull thickness varies across its surface, it acts as a random phase aberration to the transmitted waves, thus making constructive interferences less probable.

Clement et al. [33] compute the amplitudes of the reflected wave, the transmitted longitudinal wave, and the transmitted shear wave, with respect to the incident wave amplitude using the fact that the normal component of the particle displacement, the normal stress and the shear stress must be continuous at the boundary.

In White et al. [42], the skull interfaces are supposed to be parallel and the pressure (longitudinal and shear) transmission coefficients are computed with:

$$\begin{cases} T_L = \dfrac{\rho_f}{\rho_S} \dfrac{2Z_L \cos(2\theta_S)}{Z_L \cos^2(2\theta_S) + Z_S \sin^2(2\theta_S) + Z_f} \\ T_S = -\dfrac{\rho_f}{\rho_S} \dfrac{2Z_S \cos(2\theta_S)}{Z_L \cos^2(2\theta_S) + Z_S \sin^2(2\theta_S) + Z_f} \end{cases} \tag{36}$$

Where $\rho_f$ and $\rho_S$ are the densities of the fluid and the solid, $Z_L$ and $Z_S$ are the longitudinal and shear impedances of the solid, $Z_f$ is the fluid impedance. The ultrasound beam is approximated as a one-dimensional ray. The values obtained after the first interface are then used as the input for the second interface.

Yin and Hynynen [34] divide each skull interface into small rectangular planar patches. The plane-wave theory is then applied on each patch and the contribution of each patch is then summed using the Rayleigh-Sommerfeld integral. The computations are made for each interface successively: once the computations are obtained for an interface, this interface is considered as the new source for the next interface. As the patch surfaces are small (about a quarter wavelength), the acoustic pressure in tissue can be computed as the product of the particle normal velocity and the specific acoustic impedance of the tissue.

Pichardo et al. [35] also divide the interfaces into small elements, but the amplitude computations are different. The contribution coefficients due to the longitudinal propagation from the transducer surface to the outer face of the skull $U_{LL,1\to 2}$, from the outer face to the inner face $U_{LL,2\to 3}$ and from the inner face to a point $q$ inside the brain $U_{LL,3}$ are computed with:

$$\begin{cases} U_{LL,m\to n} = \dfrac{jk^*_{l_m}\rho_m c_{l_m}}{2\pi}\dfrac{e^{-jk^*_{l_m}R_{i_m\to i_n}}}{R_{i_m\to i_n}}\left(1-\dfrac{j}{k^*_{l_m}R_{i_m\to i_n}}\right)T_{l_{i_m\to i_n}}\cos\left(\theta_{i_m\to i_n}\right)ds_{i_m} \\ U_{LL,3} = \dfrac{jk^*_{l_3}\rho_3 c_{l_3}}{2\pi}\dfrac{e^{-jk^*_{l_3}R_{i_3\to q}}}{R_{i_3\to q}}ds_{i_3} \end{cases} \quad (37)$$

Where $c_{l_m}$, and $k^*_{l_m} = k_{l_m} + j\,\alpha_{l_m}$ are the longitudinal speed of sound and wave number ($\alpha_{l_m}$being the attenuation coefficient), $\rho_m$is the density, $R_{i_m\to i_n}$is the distance between the surface elements $i_m$and $i_n$, $\theta_{i_m\to i_n}$ is the incident angle, $ds_{i_m}$is the surface area of element $i_m$, and $T_{l_{i_m\to i_n}}$is the transmission coefficient. Similarly, the contribution coefficients due to shear-mode conversion are computed. The pressure amplitude of the transmitted wave is then obtained using the Rayleigh-Sommerfeld integral for a multilayer case:

$$p(q) = p_{LL}(q) + p_{SL}(q) \quad (38)$$

With $p_{LL}(q) = \sum_{i_1=1}^{N_1} u_{i_1} U_{LL,1\to 2}\, U_{LL,2\to 3} U_{LL,3}$ and $p_{SL}(q) = \sum_{i_1=1}^{N_1} u_{i_1} U_{LS,1\to 2}\, U_{LS,2\to 3} U_{LL,3}$, where $N_1, N_2, N_3$ correspond to the number of element of each interface (transducer, outer and inner surfaces of the skull) and $u_{i_1}$is the amplitude of the particle displacement of the element $i_1$.

Pajek and Hynynen [38] use the same method but shear waves are not taken into account due to their high attenuation at the high frequencies used (higher than 1MHz). Jones et al. [39] also use this method, but with spatially averaged acoustical properties for each ray path going through the skull.

In Jiang et al. [52], the frequency deviation caused by the acoustic attenuation is calculated in the frequency domain and the ultrasound amplitude after refraction is computed using the transmission coefficient at each interface. After the first interface, the amplitude of the velocity potential is given by:

$$Q_2 = Q_s\,T_r\,\frac{l_2'}{l_2' + l_{2r}} \quad (39)$$

Where $Q_s$ is the velocity potential of the source, $T_r$ is the plane-wave transmission coefficient, $l_1$ is the length of the incident ray, $l_{2r}$ is the length of the refracted ray, and $l_2' = \frac{l_1 sin\alpha\cos^2\beta}{sin\beta\cos^2\alpha}$ represents the divergence of the beam. Then after propagation through the medium layer, the velocity potential at the second interface is considered as a new source to compute the velocity potential after crossing the second interface.

### 3.1.2 Transfer function methods

#### 3.1.2.1 Transfer matrix method for transmission coefficient computation

In Pichardo et al. [37], as the skull is divided into N layers and is surrounded by water, the transmission coefficient $T$, at normal incidence, is computed using:

$$T = \frac{2Z_w}{(m_{2,2} + m_{3,3} + Z_w m_{2,3})Z_w + m_{3,2}} \quad (40)$$

Where $Z_w = \rho_w c_w$ is the acoustic impedance of water, $m_{j,k} = A_{j,k} - \frac{A_{j,1}A_{4,k}}{A_{4,1}}$($A_{j,k}$is the coefficient of the transfer matrix $A$ giving the stress and particle velocity at the last interface from the stress and particle velocity at the first interface. $A$ is obtained by the product of the matrices of the sound transmission coefficients of the $N$ skull layers).

#### 3.1.2.2 Projection algorithm in the wavevector-frequency domain

Clement and Hynynen [32] propose a projection algorithm in the wavevector-frequency domain that works for normal incidence or any angle of incidence. By taking the Fourier transform of the linearized Stoke's equation, they obtain a simple equation whose solution is given by:

$$\tilde{p}(\tilde{k}, z) = \tilde{p}(\tilde{k}, z_0)e^{i\tilde{k}(z-z_0)} \quad (41)$$

Thus, the field recorded in a plane $z = z_0$ is related to the field at any other plane by a transfer function. The pressure field is then obtained by taking the inverse Fourier transform. For non-normal incidence, a rotation of the planar field at $z = z_0$ is performed in the wavevector-frequency domain.

#### 3.1.2.3 Hybrid Angular Spectrum (HAS) method

The traditional angular spectrum method assumes that the medium is homogeneous. The pressure in the initial plane is encoded into a spectrum of traveling plane waves in the spatial-frequency domain using Fast Fourier Transform. Then, the propagation of the waves to the next plane is calculated in the spatial-frequency domain by multiplying the initial spectrum by a propagation transfer function. Finally, IFFT is performed to obtain the pressure in the final plane.

Vyas and Christensen [70] extend this method to calculate linear ultrasound wave propagation in inhomogeneous tissue geometries. In this new method, wave propagation through each plane is calculated in two steps: one in the space domain and one in the spatial-frequency domain.

The transmission from one plane to another is computed using:

$$p_n(x, y) = p_{n-1}(x, y)\ e^{jb_n(x,y)r' - a_n(x,y)r} \quad (42)$$

Where $b_n(x, y) = \frac{2\pi f}{c_n(x,y)}$ is the propagation constant at a given voxel $(x, y)$ with a speed of sound $c_n(x, y)$, and $a_n(x, y)$ is the pressure attenuation coefficient of a given voxel. $r'$ is the effective path length between two wavefronts and $r$ is the entire propagation path.

By dividing the phase change into an average phase shift $b_n' r'$ inside a plane and the difference $\Delta b_n(x, y)r'$ from this average phase shift for each voxel of the plane, the transmission equation can be re-written as:

$$p_n(x, y) = p_{n-1}(x, y)\ e^{jb_n' r' + j\Delta b_n(x,y)r' - a_n(x,y)r} \quad (43)$$

Propagation from one plane to another then is achieved in the space domain using:

$$p'_n(x,y) = p_{n-1}(x,y)\, e^{j\Delta b_n(x,y)r' - a_n(x,y)r} \tag{44}$$

And propagation across a given plane is performed in the spatial-frequency domain using:

$$p_n(x,y) = F^{-1}\left[F[p'_n(x,y)]e^{jb'_n\sqrt{1-\alpha^2-\beta^2}\,\Delta z}\right] \tag{45}$$

With $\alpha = \lambda f_x$ and $\beta = \lambda f_y$.

This method is used in [25,47,55].

## 3.2 Numerical methods

### 3.2.1 Finite differences

Many studies[6,10,28,30,37,45,46,50,53,57,59,65,13,67,71,72,14,17–21,26] use finite difference methods to simulate wave propagation. This allows modelling the skull as a heterogeneous medium but it is computationally intensive. Unlike semi-analytical methods, finite differences methods need a spatial step fine enough in order to converge, as the pressure field at a given position depends on the pressure field at previous positions.

Several wave equations can be considered when using finite differences, depending on which effects one wants to take into account or neglect, such as linearity, shear mode conversion, absorption etc. All those equations are derivatives of the Westervelt equation, which is given by:

$$\frac{\partial^2 p}{\partial z^2} - \frac{1}{c^2}\frac{\partial^2 p}{\partial t^2} + \frac{\mu}{c^4}\frac{\partial^3 p}{\partial t^3} + \frac{2\beta}{\rho c^4}\left[p\frac{\partial^2 p}{\partial t^2} + \left(\frac{\partial p}{\partial t}\right)^2\right] - \frac{\partial p}{\partial z}.p\frac{\partial(\ln p)}{\partial z} = 0 \tag{46}$$

Where $p$ is the wave pressure, $c$ is the local speed of sound, $\rho$ is the local density, $\mu = 2\,\alpha c^3(2\pi f)^{-2}$ is the local diffusivity parameter ($\alpha$ is the local attenuation coefficient) and $\beta$ is the non-linearity parameter.

*Table 1: Effects taken into account in the equations used in finite differences studies*

| Reference | Non-linearity | Shear waves | Absorption |
|---|---|---|---|
| 20,26,29,30,37 | X | | X |
| 6,10,71,13,18,20,28,45,46,53,57 | | | X |
| 21 | | | |
| 59 | | X | |
| 14 | | X | X |
| 17 | X | X | X |

### 3.2.2 k-space

In 2012, Jing et al. [51] propose to use a k-space method to simulate wave propagation inside the skull. Like finite differences, it allows a heterogeneous description of the skull, but k-space is faster as it needs less spatial resolution to achieve convergence with a reasonable accuracy.

The k-space methods are part of the bigger family of Pseudo-Spectral Finite Difference Time Domain methods (PSTD). The PSTD methods transform the fields to the spectral domain at each time step in order to compute the spatial derivatives more easily. Those methods are called global methods because the entire simulation domain is used in approximating the derivative of a single point. Those methods are therefore more accurate than standard finite difference methods, and theoretically allow discretizing the domain as low as two points per wavelength. However, the standard PSTD methods only improve the efficiency in the spatial domain, as finite differences schemes are still needed in the time domain, and thus fine time steps are needed to avoid dispersion. The k-space method aims to allow larger time steps without lowering the accuracy too much. It consists in multiplying the time step by the k-space operator $\kappa = sinc(c_{ref} k \frac{\Delta t}{2})$, which is derived from the exact solution of the homogeneous and lossless wave equation.

The k-space method has been used in other studies [5,15,23,24,49,54,66–69], thanks to the development of the Matlab toolbox k-wave, even if some studies have developed their own k-space algorithm [52].

### 3.3 Hybrid methods

A good balance is to combine numerical methods and analytical ones. For instance, one can use an analytical method to compute the wave propagation before and after the skull (as those media can reasonably be assumed homogeneous) and then use a numerical method only within the skull.

When simulations are used for phase correction computations, the waves are propagating from a virtual source inside the brain towards the transducer outside the skull. In a few papers [53,54,71,73,74], the finite difference or k-space algorithms are only performed from an arbitrary plane or spherical surface as close as possible to the inner surface of the skull, and propagation in brain tissues is modelled by a ray-tracing code to save time. In [50,54,74], the numerical simulations are only performed to an arbitrary plane or a receiving surface just in front of the skull, and the propagation from there to the transducer is realised by a ray tracing code.

In direct simulations (from the transducer to the brain), hybrid methods are also used. In Wu et al. [68], they developed a method that combines the k-space corrected PSTD method with an acoustic holography approach based on the Rayleigh integral. The k-wave stage is used to calculate the sound field in the skull, which is divided into multiple sound paths (one per array element). The propagation in each sound path is simulated with k-wave from the element to a virtual sensor located near the skull-brain interface. The simulations in all sound paths are run in parallel. The relative amplitude and phase of each discrete point in the sensors are extracted by taking an FFT of an integer number of cycles of the signal after it reaches steady state. Propagation from the virtual sensors is then performed with the Rayleigh integral based method. In Pulkkinen et al. [27], the propagation from the phased-array element through the water to a parallel plane near the skull is simulated with the Rayleigh integral. Then the propagation through the skull is performed with a finite element code and propagation in the brain is simulated using the angular spectrum method.

### 3.4 Comparison of the methods

All these methods simulate wave propagation through the skull with a variable degree of realism, depending on the acoustic effects taken into account or neglected. Most of the time analytical methods are faster but less accurate than numerical ones. A few papers have compared some of these methods.

In Jing et al. [51], they compared the k-space method and the Finite Differences Time Domain (FDTD) method. It was found that for very fine spatial resolution (more than 10 grid points per wavelength), these two methods match very well. However, at a low spatial resolution, the k-space method was observed to produce considerably less numerical error. The computation time at a fine spatial resolution (7.68 grid points per wavelength) can be reduced by a factor of around 80 using 2.56 grid points per wavelength, without altering too much the resulting field.

Jones et al. [6] compare full-wave finite difference phase correction with analytical phase correction. The results show that the finite difference method outperforms the analytical one for every metrics (shift decrease of 25%, -3dB volume decrease of 20%, peak-to-side-lobe ratio increase of 26%, signal-to-noise ratio increase of 37% with the full-wave method compared to the analytical one). However, the FDTD method took 7.8h while the analytical phase correction took only 27s.

Kyriakou et al. [20] compare distance based phase correction, ray-tracing based phase correction and finite difference based phase correction. The ray-tracing based phase correction only performs a bit better than the distance based one, and gives much worse results than the finite difference based phase correction (shift x6, peak pressure /2, focal volume x5).

Robertson et al. [67] compare FDTD and k-space methods for phase aberration correction. Numerical dispersion has a serious effect on the accuracy of FDTD scheme, resulting in high temporal sampling requirements to reduce positional error, while for the k-space scheme, only 3 Pixel Per Wavelength (PPW) will serve to limit dispersion sufficiently for transcranial transmission for any stable Courant-Friedrichs-Lewy (CFL) value. To reduce error in the intensity below 10% following transcranial transmission, k-space scheme requires 4.3 PPW, while FDTD requires 5.9 PPW.

Jiang et al. [52] compare FDTD, k-space and ray-tracing methods, for phase and amplitude correction. The k-space method (4 PPW) has a 0.7% phase error compared with the FDTD method (16 PPW). The ray-tracing method has an amplitude error of 5.35% and a phase error of 1.2% compared to the FDTD method. The k-space method took 23h35 while the ray-tracing method took 37min.

Bancel et al. [54] compare k-space and ray-tracing methods. The restored pressure compared to the hydrophone-based correction is 85% for the k-space method and 83.5% for the ray-tracing method. Similarly, the other metrics show results a bit better with the k-space method but the differences are not as big as other studies have shown before.

In terms of computation speed, it is quite hard to compare the methods for various papers as the conditions (CPUs, GPUs, parallel computing, …) differ from one paper to another and relevant information (such as domain size, time step) are not always available.

*Table 2: Comparison of the computation time of several simulation methods*

| Ref | Simulation method | Skull modelling | Computer | Domain size (mm3) | Number of points | Time | Time for 1e9 points |
|---|---|---|---|---|---|---|---|
| [32] | Layered wavenumber | Homogeneous | 1GHz | | | 5h | |
| [18] | Finite differences | Heterogeneous | 500MHz | 70x10x30 | 2,1e7 | 20h | 950h |
| [33] | Ray-tracing | Locally homogeneous | 1GHz | | 1,6e4 | 30s | 521h |
| 35 | Ray-tracing | Homogeneous | | 160x160x160 | 3,2e7 | 106min | 55h |
| 21 | Finite differences | Heterogeneous | | 180x180x150 | 1,5e9 | 90min | 60min |
| 65 | Finite differences | Heterogeneous | | | 2,2e9 | 120min | 54min |
| 30 | Finite differences | Binary | 2,8GHz | | 1,0e8 | 12h | 120h |
| 72 | Finite differences | Heterogeneous | | | 1,3e9 | 3h | 2h18 |
| 38 | Ray-tracing | Homogeneous per ray | | | | 24h | |
| 51 | k-space | Heterogeneous | 2,6GHz | 47x100x31 | 7,5e5 | 17s | 6h20 |
| 6 | Finite differences | Heterogeneous/Homo | | 100x130x40 | 2,1e7 | 7h48 | 371h |
| 55 | HAS | Heterogeneous | 2,7GHz | | | 20min | |
| 15 | k-space | Heterogeneous | 2,5GHz | | 1,9e6 | 15min | 131h30 |
| 47 | HAS | Heterogeneous | | | 4,6e7 | 45min | 16h |
| 11 | k-space | Heterogeneous | | | 1,2e8 | 30h | 250h |
| 75 | k-space | Heterogeneous | 3,3GHz | | 2,7e8 | 19h53 | 74h |
| 69 | k-space | Binary | 3,3GHz | | 4,2e6 | 9h | 2143h |
| 57 | Finite differences (GPU) | Heterogeneous | 3,6GHz | 180x200x175 | 6,3e7 | 150s | 40min |
| 25 | HAS (7 CPU cores) | Heterogeneous | 1,4GHz | | 1,4e9 | 180s | 128s |
| 52 | k-space | Heterogeneous | | | 1,3e8 | 23h | 177h |
| 52 | Ray-tracing | Homogeneous | | | 1,3e8 | 37min | 4h40 |

## 4. Experimental validation of the simulations

Experiments are the only way to evaluate the accuracy of simulations. They are most of the time considered as gold standards or references even though they can also be enticed with errors as discussed below. In this section, the classical set-ups for experimental validation of transcranial acoustic simulations are described and the experiments versus simulations results are compared.

### 4.1 Experimental set-ups

The typical set-up for transcranial acoustic measurements is composed of a transducer and an hydrophone mounted on a three axis positioning system. A first scan, without skull is often performed as a reference result. Then, a skull or a phantom is placed between the transducer and the hydrophone. Prior to the experiments, the skulls are put in water and degassed (typically for 24h) in order to remove all the air trapped inside and to ensure the skull pores are filled with water to mimic *in vivo* conditions. When the skull modelling is deduced from CT scan, CT-scans of the skulls should be performed under water as well. Likewise, the experiments are made in a tank filled with degassed water.

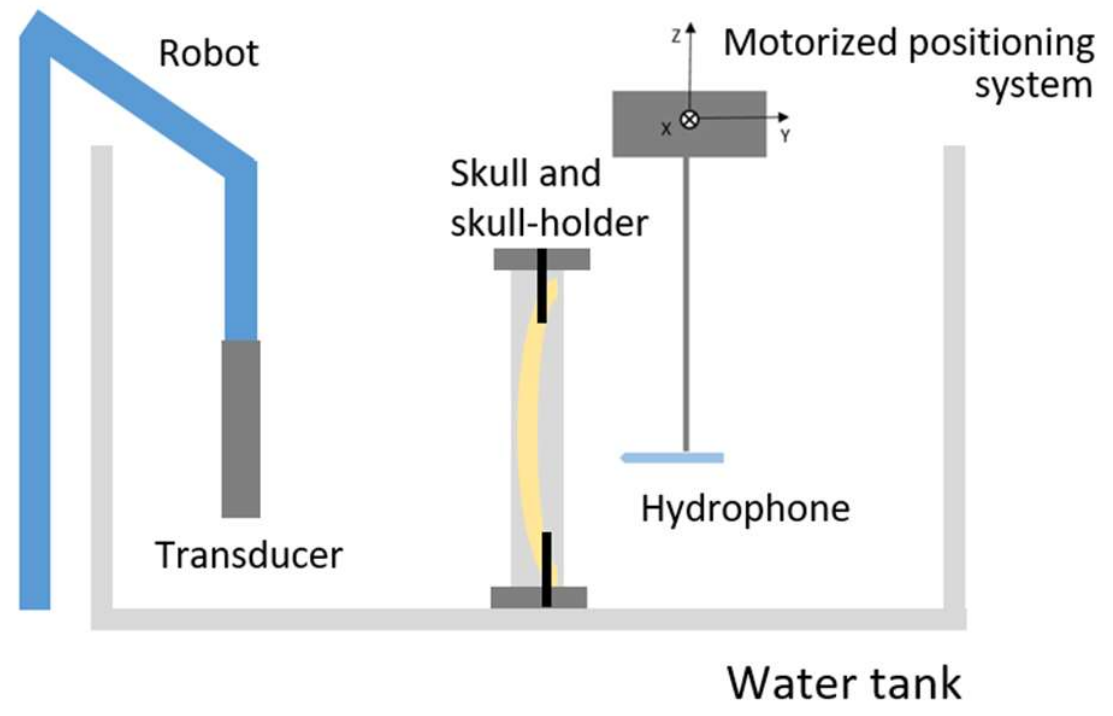

*Table 3: Typical set-up for transcranial acoustic measurements*

Most of the time, accurate positioning of the skull is performed using a stereotactic frame, or a homemade frame, which contains fiducial markers. The CT scans of the skulls are acquired with the positioning frame and the position of the fiducial markers is determined from the CT images. Likewise, the fiducial marker positions are recorded with the three axis positioning system [18,46], or with an optical-based position tracking system [57]. From this, a rigid transformation matrix representing the rotation and translation between the experimental and simulation frames can be computed [32]. Otherwise, the transducer can be attached to the skull positioning frame [5,32,37,50,68,71,74] in order to obtain accurate relative positioning more easily. On the other hand, this method does not allow for different positions of the transducer. When experiments are performed under MR guidance, MR tracking coils can be used to find the transducer location within the MR frame [24]. MR images can also be used to find the relative position of the transducer with respect to the skull frame [47].

A few studies have estimated the positioning error between their experiments and their simulations. Those errors are displayed in the following table:

*Table 3: Positioning method and error for measured made in several studies*

| Reference | Positioning method | Positioning error |
|---|---|---|
| [32] | Skull attached to the transducer | 0.5mm |
| [18] | Position of the skull markers recorded by the three axis system | 1mm |
| [50] | Skull attached to the transducer | A few millimetres |
| [71] | Skull attached to the transducer | 0.7mm |
| [47] | Relative position between transducer and skull measured with MRI | 0.25mm |
| [57] | Position of the skull markers recorded by an optical based position tracking | 0.41mm |
| [68] | Skull attached to the transducer | 0.02mm |

O'Reilly et al. [76] use high frequency ultrasound measurements to localize the skull surface and register CT data to the ultrasound treatment space. The results show on an average sub-millimetre (0.9±0.2 mm) displacement and sub-degree (0.8◦±0.4◦) rotation registration errors.

### 4.2 Performance vs simulation

Though many studies have checked their simulation results through experiments, only a few of them have directly compared the pressure field obtained with experiments with the simulated one. Indeed, most studies have compared simulations and experiments through their application. For instance, many studies experimentally compared the field produced by a multi-element transducer with a simulated phase correction with the field produced by a multi-element transducer with a hydrophone based correction (which is considered as the optimal phase correction) [5,17,18,32,47,53,54,68,71,74]. Another way to assess the accuracy of transcranial acoustic simulations is to compare the results of heat simulations with MR thermometry data [25] or with lesion locations [24].

In Bouchoux et al. [46], Yoon et al. [57] and Bancel et al. [54], they directly compare the acoustic field obtained with experiments with the simulated one. They obtain the following results:

*Table 4: Comparison between experimental and simulated acoustic field in different studies*

| Reference | Focus shift | Pressure metric | Pressure value |
|---|---|---|---|
| 46 | Below the scanning step (2.5mm) | $\sum_{x,y,z} \frac{\max_t \lvert p_{simu} \rvert - \max_t \lvert p_{meas} \rvert}{\max_t \lvert p_{meas} \rvert}$ | -9.17 ± 4.55%, averaged on 4 skulls (120kHz) |
| 57 | 1.43 ± 0.8 mm | $\frac{R^{XY} - \overline{R^{XY}}}{R^{XY}} + \frac{R^{XZ} - \overline{R^{XZ}}}{R^{XZ}} + \frac{R^{ZY} - \overline{R^{ZY}}}{R^{ZY}}$ With R the ratio of the peak pressure with and without the skull. | 5.8 ± 6.4 %, averaged on 3 sheep skulls (250kHz) |
| 54 | 0.3 ± 0.1 mm | $\frac{1}{\mathrm{N_{elements}}} \sum_{i=1}^{N_{elements}} p_i^{simu} - p_i^{meas}$ | -0.9 ± 5.4%, averaged on 5 skulls (800kHz) |

The phase correction results obtained in several studies are displayed in the next table:

*Table 5: Comparison of simulation-based and invasive aberration correction in several studies*

| Reference | Shift | | | Percentage of peak pressure compared to the uncorrected case | |
|---|---|---|---|---|---|
| | Without correction | With simulation-based correction | With hydrophone-based correction | With simulation-based correction | With hydrophone-based correction |
| 32 | 1.1mm | 0.48mm | | 135% | 294% |
| 17 | | 0mm | | | |
| 47 | 1.77mm | 0.71mm | 0.25mm | 151% | 217% |
| 53 | 1.05mm | 0.54mm | | 190% | 220% |
| 68 | 0.5mm | 0mm | | 137% | 147% |
| 54 | 0.9mm | 0.2mm | 0mm | 130% | 153% |
| 74 | | 0.63mm | | 149% | 165% |

Almquist et al. [47] perform experiments with phantoms and human skulls in order to separate the errors coming from the simulation algorithm from the errors coming from the skull modelling (as the acoustical properties of the phantoms are known). The results are presented below:

*Table 6: Comparison of simulated-based and invasive aberration correction for a phantom and for a human skull*

| Metric | Phantom | | | Skull | | |
|---|---|---|---|---|---|---|
| | Without correction | With simulation-based correction | With hydrophone-based correction | Without correction | With simulation-based correction | With hydrophone-based correction |
| Shift | 1.9mm | 0.56mm | 0.25mm | 1.77mm | 0.71mm | 0.25mm |
| Percentage of peak pressure compared to the uncorrected case | 100% | 141% | 150% | 100% | 151% | 217% |

Simulation based correction is able to reach 94% of the optimal correction (hydrophone-based correction) when the acoustical properties are known (phantom experiments) while only 70% when they are unknown and deduced from CT-scans with empirical relationships (skull experiments).

In a few studies [23,25–27], MR thermometry data was compared with the results of thermal simulations. Qualitatively speaking, the predicted and measured temperature fields agree well in all those studies. In Pulkkinen et al. [27], for the seven sonications performed on a single skull, the error between the simulated and measured temperatures at the two locations of interest (focus and tissue near the skull base) is on average 1°C and never exceeds 3°C. The other studies collected data from several patients (at least 9) who had undergone several sonications, and were thus able to make statistical analysis of the results.

*Table 7: Correlation between simulated and measured temperature in several studies*

| Reference | Metric | Correlation between simulated and measured |
|---|---|---|
| 23 | Peak temperature rise | $r^2 = 0.71$ |
| 26 | Peak temperature | $r^2 = 0.75$ |
| 25 | Focal spot temperature rise | $0.613 \leq r^2 \leq 0.947$ |

McDannold et al. [23] also compare the dimensions and obliquity of the heating. The found that the predicted and measured dimensions ($R^2 = 0.62$) and obliquity ($R^2 = 0.74$) of the heating are correlated.

### 4.3 Reasons for error between experiment and simulation

On the simulation side, as shown by Almquist et al. [47], imprecise skull modelling based on CT scans is an important source of mismatch between simulations and experiments. In addition, many approximations, such as neglecting shear waves (especially if the incidence angles are greater than 30° [33]), attenuation and non-linearity, degrade the accuracy of the results. For instance, in Ding et al. study [29], the thermal dose at the focus computed with non-linear wave propagation was almost twice the linear thermal dose, suggesting that non-linearity effects have to be taken into account. Jiang et al. [52] compare simulations with and without shear waves for incident angles smaller than 20°. They found a shift of 0.5mm of the focus when shear waves are neglected, while perfect focusing is reached when shear waves are included. The maximum pressure reached with shear waves is approximately 3.65 Pa, while that when shear waves are not considered is approximately 4.85 Pa.

On the experimental side, very close positioning of the experiments relatively to the simulations is crucial in order to be sure to be able to compare them. One also needs to keep in mind that hydrophones are generally not very precise (up to 10% errors) and need to be well calibrated.

## 5. Discussion

Almquist et al. [47] showed that transcranial acoustic simulations accuracy is highly dependent on skull modelling, which can be divided into two steps: the geometric description and the acoustic properties. The shape is not needed for numerical methods and is usually segmented from CT or MR images for semi-analytical methods. Over the years, image-derived acoustic properties have replaced measured ones, because of their case-specific adaptability. Most studies agree that both density and speed of sound depend linearly on HU values, but various empirical relationships have been proposed and improved. Marsac et al. [53] seem to have reached the more general porosity-based relationships and their accuracy is demonstrated by Bancel et al. [54]. The main issue with these relationships remains to choose the constants ($\rho_{max}, \rho_{min}, c_{max}, c_{min}$). While $\rho_{min}$ and $c_{min}$ are often taken as the values of water, $\rho_{max}$ and $c_{max}$ are sample-dependant and thus hard to choose. Marsac et al. [53] tried to find the best $c_{max}$ such that simulations fit the experiments, but this kind of approach is hard to generalize as it depends mostly on the samples used, the frequency and the experimental conditions. Indeed, Pichardo et al. [37] showed that the speed of sound relationships are frequency dependent. A polynomial relationship has been proposed by McDannold et al. [23] for the speed of sound, with constants found by comparing simulations with experiments. Once again, the problem of such strategies is that the relationships are case specific to the experimental conditions of the study and are not always well generalizable. To avoid any bias, the relationships need to be compared in a different study, such as the one from Bancel et al. [54], who showed that Marsac relationships were more accurate than Pichardo's and McDannold's. But more comparison studies are needed. Finally, attenuation is the parameter that is the hardest to determine since it is not well understood. Various empirical relationships have been proposed, relating attenuation to HU values with a linear or a power law. Robertson et al. [49] showed that attenuation can be defined by a constant across the whole skull, but of course this constant is skull specific. Linear and power laws relating attenuation with frequency have also been investigated. In particular, in trabecular bone, where most of the scattering occurs, linear relationships have been established with a slope that varies quasi linearly with the bone fraction volume. However, Yousefian et al. [61] claim that total attenuation is not described by a linear combination of scattering and absorption anymore in the presence of multiple scattering. They propose a relationship for attenuation, which is the sum of a constant absorption and a frequency dependent power law for scattering. However, to our knowledge, no comparison studies have shown that any attenuation relationship is better than the others. While most studies derived the acoustic properties of the skull from CT scans, Webb et al. [64] warned that X-ray energy, reconstruction method, and reconstruction kernel should also be taken into account. After computing the acoustic properties of the skull, one can consider several levels of heterogeneity, going from global properties for the whole skull to individual properties for each voxel. Comparisons of these levels of heterogeneity is complicated as it is closely linked to the kind of simulation method used. Indeed, numerical methods tend to use fully heterogeneous skull model while semi-analytical methods (in particular ray-tracing) often use fully homogeneous ones. Jones and Hynynen [6] and Kyriakou et al. [20] both compared homogeneous and heterogeneous models with a same simulation method for correction aberration and found that, as can be expected, the heterogeneous performs better, especially in terms of predicted pressure amplitude, while both models seem nearly equivalent in terms of predicted shift. This is in agreement with Robertson et al. study [69], who show that homogenization of acoustic property maps leads to an overestimation in the amplitude of ultrasound and an underestimation of time-of-flight. However, smarter homogeneous models can be used to improve computation time without losing too much accuracy. For example, Jiang et al. [77] showed that taking different averaged properties for the refraction computations and for the time-of-flight computations significantly improves the accuracy without increasing the computation time.

In terms of simulation methods, the most used ones are ray-tracing, Hybrid Angular Spectrum, FDTD and k-space. The k-space method seems to have replaced FDTD, as it produces less numerical error at low spatial resolution [51]. Thus, the k-space method can be used with quite coarse meshes without losing to much accuracy, which results in a computation time decrease (by a factor of around 80 [51]). Several studies compare finite difference based [6 20] and k-space based [52 54] phase correction with ray-tracing based phase correction. The results show that both FDTD and k-space based phase correction outperforms ray-tracing based phase correction, but the gap between numerical methods and ray-tracing vary between the studies (going from a 2% difference [54] to a 50% difference [20] in restored pressure at focus). In terms of computation time, FDTD is slower than ray-tracing by three orders of magnitude Jones et al. [6], while the k-space method is slower than ray-tracing by one order of magnitude [52].

6. Conclusion

Transcranial focused ultrasound is a promising method for several therapeutic applications. Simulations are needed to optimize such treatments and to assess their safety. The main challenge of such simulations is to determine the skull acoustic properties. Then, depending on level of precision of the skull modelling, different types of simulation methods can be used. Another big challenge is to make measurements precise enough to be able to compare them with simulation results and thereby to assess the accuracy of the simulations.

The authors have no conflict to disclose.